\documentclass{article}
\usepackage{graphicx}
\usepackage{amsmath}
\usepackage{amsfonts}
\usepackage{amssymb}

\begin{document}

\title{Uniqueness of Inverse Scattering Problem in Local Quantum Physics\\{\small dedicated to Moyses Nussenzveig on the occasion of his 70th birthday}}
\author{Bert Schroer\\present address: CBPF, Rua Dr. Xavier Sigaud 150, \\22290-180 Rio de Janeiro, Brazil\\email schroer@cbpf.br\\permanent address: Institut f\"{u}r Theoretische Physik\\FU-Berlin, Arnimallee 14, 14195 Berlin, Germany}
\date{February 2003}
\maketitle
\begin{abstract}
It is shown that the operator algebraic setting of local quantum physics leads
to a uniqueness proof for the inverse scattering problem. The important
mathematical tool is the thermal KMS aspect of wedge-localized operator
algebras and its strengthening by the requirement of crossing symmetry for
generalized formfactors. The theorem extends properties which were previously
seen in d=1+1 factorizing models. 
\end{abstract}

\section{Inverse Problem in LQP}

Most inverse problems have their origin in classical physics where they result
from the question to what extend scattering (asymptotic) data allow a
reconstruction of local data. The universality and importance of this kind of
quest inspired Marc Kac to the famous aphorism ``How to hear the shape of a
drum'' which refers to Weyl's problem associated with geometric
reconstructions from the asymptotic distribution of eigenvalues of the
Laplace-Beltrami operator. In its present use it incorporates a wide range of
problems concerning the partial or complete determination of local data from a
seemingly weaker asymptotic input. In local quantum physics also the problem
of how to re-construct the full content of QFT (in particular the
superselection aspects of charge-carrying fields) from its ``observable
shadow'' i.e. the so-called DR-theory \cite{DR} may be viewed as an inverse problem.

The inverse problem in the present context is the question to what extends the
data contained in a physically admissible (unitary, crossing symmetry,...)
S-matrix determines a quantum field theory. Since it is easy to see that an
S-matrix cannot be uniquely related to one field but rather is shared by very
\ big equivalence classes of local fields, it is clear that the first step in
such an investigation is to formulate QFT in a way that two isomorphic
theories whose different appearance is only due to the use of different
``field-coordinatizations'' are easily recognizable as being one and the same.
This is not possible (or rather extremely cumbersome) in the standard approach
based on pointlike fields.

Fortunately there exists such a framework which pays due attention to fields
belonging to the same local equivalence class (fields from the same Borchers
class \cite{St-Wi}\cite{Haag}) and which therefore generate the same system of
local algebras as the given field. It is known under two different names
algebraic quantum field theory (AQFT) or local quantum physics (LQP)
\cite{Haag}; in fact one of the motivations behind its creation was precisely
a better understanding of the insensitivity of the S-matrix against certain
kind of local changes of field-coordinatizations. Its relation to standard QFT
is similar to that of coordinate-based differential geometry to its more
modern coordinate-free intrinsic formulation. Here as there one retains all
the underlying principles and only introduces additional more elegant concepts
to implement them. If one wants to emphasize that one deals with QFT but
without the reader immediately thinking in terms of Lagrangians and functional
integral representations it is quite helpful to use AQFT or LQP instead of QFT
(a name as ``intrinsic QFT'' or ``field-coordinatization-free QFT'' would
appear a bit clumsy).

Using the powerful mathematical tool of Tomita-Takesaki theory adapted to
wedge algebras \cite{Bo} as well as the closely related crossing symmetry, we
show that if a solution exists at all, it is necessarily unique.

This result generalizes previous special findings in d=1+1 models (for a
review in the LSZ setting of scattering theory see \cite{BFKZ}) concerning the
modular derivation of the wedge localization interpretation \cite{S1}\cite{S2}
of the Zamolodchikov-Faddeev algebra \cite{Zam} in d=1+1 in which case one
also is able to control a good part of the explicit construction of the
associated QFT. \ For such models Zamolodchikov and Faddeev observed that the
structure of their factorizing S-matrix can be encoded into an interesting
algebraic structure. Although the objects which generate this algebra were
left without a physical interpretation, they seemed to constitute a quite
useful technical tool for the construction of generalized formfactors (matrix
elements of operators between physical multi-particle state vectors)
affiliated with the model S-matrix \cite{Smir}. The application of modular
theory to this situation revealed that the rapidity dependent Z-F operators
are special cases of ``(vacuum) \textbf{p}olarization-\textbf{f}%
ree-\textbf{g}enerators'' (PFG) for the wedge-localized algebra \cite{S1}%
\cite{S2}. In this special setting of ''tempered PFGs'', the relation between
the crossing symmetry in momentum space as formulated in the LSZ theory and
the KMS thermal condition following from Bisognano-Wichmann properties of
wedge algebras is well-understood\footnote{For the case of absense of bound
states, a detailed and mathematically rigorous use of the modular theory
(which takes into account the domain problems of unbounded operators
affiliated with $\mathcal{A}(W))$ can be found in a recent paper \cite{Lech}.}.

In the general case the PFG's have unwieldy domain properties \cite{BBS} which
prevent their use in relating momentum space crossing with the spacetime KMS
property of wedge localization. In this case we will separately assume the
validity of the crossing property and show that if we combine this with the
deduced thermal KMS\ for wedge algebras we may obtain the desired uniqueness
of the LQP (without being presently able to say something about the
existence). It should not be a surprise that with our assumptions being
stronger than those used in \cite{Bu-Fre} we also get a stronger statement.
For example we exclude in case of S=1 any would be solution beyond
Wick-polynomials or in a more algebraic language beyond the free Borchers-class.

\section{The use of modular theory in wedge localization}

The new tool on which the proof relies in an essential way is the modular
theory of operator algebras as discovered and elaborated by Tomita and
Takesaki\footnote{For an eyewitness account of the birth of this theory as
well as a mathematical physics presentation of its content see \cite{Bo}.}.

Since even in the setting of QFT the word ``modular'' occurs with different
meanings, we will briefly define its present use. Actually the more common use
is that of ``modular invariance'' in the more special setting of chiral
conformal field theory. Although this is not its present meaning, a future
deep connection of the T-T operator modular theory to this causality-related
classification tool of ``modular invariance'' for chiral conformal models
\cite{Itz} is by no means excluded.

Let us start from a special geometric situation encountered in QFT. Consider
the $x_{0}$-$x_{1}$ wedge $W_{0}$ in d-dimensional Minkowski spacetime which
is defined by taking convex combination of the two light rays generators
($\pm$1,1,0..0) and on which the affiliated Lorentz boost $\Lambda_{W_{0}%
}(\chi)$ acts as a wedge-preserving diffeomorphism. Imagine that a quantum
field theory is given to us and we are interested to study the structure of
the $\mathcal{A(}W_{0}\mathcal{)}$ subalgebra which the fields generate if we
limit the support of smearing functions to the wedge region (the rigorous
aspects of operator algebras will be postponed to later). The boost acts on
this wedge algebra as a unitary implemented automorphism. Let us change the
rapidity parametrization by the factor $2\pi$ and write the unitary
representation in QFT in terms of a selfadjoint $K$ boost generator as
\begin{equation}
\Delta^{it}\equiv U(\Lambda_{W_{0}}(\chi=2\pi t))=e^{2\pi itK} \label{K}%
\end{equation}
We then may introduce an unbounded positive operator $\Delta^{\frac{1}{2}}$by
``analytic continuation in t'', which in functional calculus terms means that
we are restricting the Hilbert space to those vectors $\psi$ which upon action
with $\Delta^{it}$ lead to a vector-valued function $\psi(t)\equiv\Delta
^{it}\psi$ which is continuous in $-i\pi\leq Imt\equiv\theta\leq0$ and
analytic on the open strip $-i\pi<\theta<0\footnote{Most of the analyticity
properties in x- or momentum- space rapidities result from domain properties
of unbounded operators.}.$ In addition to the wedge related boost, we also
consider the antiunitary (since it involves time reversal) reflection along
the edge
\begin{align}
J  &  \equiv U(r)\\
r  &  :x_{0},x_{1}\rightarrow-x_{0},-x_{1}\nonumber
\end{align}
which up to a $\pi$-rotation around the $e_1$-axis is identical to the famous
TCP transformation of particle physics. Since this transformation commutes
with the boost $\Delta^{it},$ its antiunitarity leads to the commutation
relation (always on the relevant domains)
\begin{align}
JK  &  =-KJ\label{J}\\
J\Delta^{\frac{1}{2}}  &  =\Delta^{-\frac{1}{2}}J\nonumber
\end{align}
on the respective domains of definition. This in turn yields
\begin{align}
S^{2}  &  \subset1\\
for\,\,S  &  \equiv J\Delta^{\frac{1}{2}}\nonumber
\end{align}
i.e. we encounter the rare case (not even to be found in extensive textbooks
on mathematical physics as that by Reed-Simon) of an unbounded antilinear
operator which is involutive and ``transparent'' on its domain i.e.
$Dom(S)=Range(S)$. In a moment we will see that it is just this somewhat
exotic property which enables the encoding of concrete spacetime geometric
information concerning quantum localization into abstract domain properties.

It is the content of a theorem (the Bisognano-Wichmann theorem \cite{Haag})
that this operator is Tomita's famous S-involution for the operator algebra
which the local fields generate if one restricts the smearing functions to
have support in the wedge. Namely in terms of the affiliated von Neumann
algebra $\mathcal{A}(W_{0})$ the $S$ fulfills the defining relation
\cite{Haag}\cite{Bo}
\begin{equation}
SA\Omega=A^{\ast}\Omega\label{equ}%
\end{equation}
where $\Omega$ is the vacuum vector, the star denotes the standard Hermitian
adjoint in operator algebras and S turns out to be closable (where the same
letter is alsso used for its closure) whose polar decomposition leads
precisely to modular objects $\Delta^{\frac{1}{2}},J$ which are related to the
unitary boost operator and the TCP related antiunitary implementer of the
reflection along the edge of the wedge. This is a special case of the Tomita
Takesaki modular theory whose prerequisite is the existence of a von Neumann
algebra in ``general position'' i.e. a pair ($\mathcal{A}$,$\Omega$) with
$\Omega$ being a cyclic and separating vector for $\mathcal{A}%
\footnote{Physicists who independently developed these concepts, often
(especially in chiral conformal field theory) talk about the unique
``operator-statevector relation'' $A\longleftrightarrow A\Omega;$ the correct
mathematical backing is the reference to the Reeh-Schlieder theorem
\cite{St-Wi}. It is often overlooked that the relation is not universal but
depends on the chosen local algebras$\mathcal{A}(\mathcal{O}).$}.$ From this
input Tomita and Takesaki derive:

\begin{itemize}
\item  The Tomita operator $S$ defined by (\ref{equ}) is a densely defined
closed antilinear involution whose polar decomposition $S=J\Delta^{\frac{1}%
{2}}$ leads to an antiunitary reflection $J$ (abstract generalization of a TCP
reflection) and modular dynamics $\Delta^{it}$ (abstract generalization of a
Hamiltonian). In contrast to our geometric approach which started from
properties of boosts and reflections along the edge of a wedge. The Tomita
setting is abstract in that $S$ is not defined geometrically but rather by
association with a ``standard pair'' ($\mathcal{A}$,$\Omega$) in operator
algebra theory. The standardness of the wedge algebra together with the vacuum
state in QFT is nothing else than a special realization of the famous
Reeh-Schlieder property in local QFT.

\item  The $J,\Delta^{it}$ have the following significance with respect to the
operator algebra
\begin{align}
AdJ\mathcal{A}  &  =\mathcal{A}^{\prime}\\
Ad\mathcal{A}  &  =\sigma_{t}(\mathcal{A})\nonumber
\end{align}
Here as in the sequel the upper dash on an operator algebra denotes its
commutant, the adjoint action of the modular unitary $\Delta^{it}$ implements
the modular group $\sigma_{t}(\cdot)\equiv Ad\Delta^{it}$ which only depends
on the state $\omega(A)=\left(  \Omega,A\Omega\right)  ,\,A\in\mathcal{A}$ and
not its implementing vector $\Omega.$

\item  A necessary and sufficient condition for the standardness
(cyclicity+separating property) of the pair ($\mathcal{A},\Omega$) is the
thermal KMS property in terms of the state $\omega$ is: there exists a $2\pi
$-open-strip analytic function (continuous in the closed strip ) $F_{A,B}(z)$
with
\begin{align*}
F_{A,B}(t)  &  \equiv\omega(\sigma_{t}(A)B)\\
\omega(B\sigma_{t}(A))  &  =lim_{z\rightarrow t+i}F_{A,B}(z)
\end{align*}
\end{itemize}

The above theorem of Bisognano and Wichmann may now be rephrased as saying
that those operator wedge algebras which are generated by covariant fields do
have a \textit{geometric} modular theory.

In more recent times there have been successful attempts to establish these
geometric modular aspects of wedge algebras directly in the seemingly more
general setting of algebraic QFT which avoids the use of fields already at the
start \cite{Bo-Yn}. The validity of this KMS condition (with the modular group
acting geometrically as the Lorentz-boost) is sufficient for establishing that
also $J$ acts geometrically i.e. that the von Neumann commutant is localized
in the geometrically opposite wedge $W^{\prime}$ (Haag duality) $\mathcal{A}%
(W)=\mathcal{A}(W^{\prime})^{\prime}$.

Starting with one standard wedge algebra $(\mathcal{A}(W_{0}),\Omega),$the
Poincar\'{e} group generates a net of wedge algebras ($\mathcal{A}(W),\Omega
$)$_{W\subset\mathcal{W}}$ and vice versa, a net of wedge algebras whose
modular data fulfill the prerequisites of the B-W theorem, generate a
Poincar\'{e} group symmetry which is uniquely determined from the modular
groups of the wedges \cite{Bo}. In order to extract sufficient physical
informations one needs nets for smaller compact causally closed regions. A net
of double cones $D$ may be defined in terms of intersections
\begin{equation}
\mathcal{A}(D)\equiv\cap_{D\subset W}\mathcal{A}(W)
\end{equation}
In order to achieve our goal we must be able to relate the wedge algebra with
the scattering operator $S_{sc}.$ This is possible in the LSZ framework of QFT
because although the representation theory of the connected Poincar\'{e}-group
for the incoming (outgoing) free fields is the same as for the interacting
Heisenberg fields, this is not so for the reflections involving time reversal.
In particular the $J$ in (\ref{J}) which represents the wedge reflection in
the presence of interactions is different from its interaction-free asymptotic
counterpart \cite{S1} $J_{in}$%
\begin{equation}
J=S_{sc}J_{in}%
\end{equation}
This implies that in the characterization of the wedge-localized (dense)
subspace:
\begin{align}
H(W)  &  =H_{R}(W)+iH_{R}(W)\\
H_{R}(W)  &  =real\text{\thinspace\thinspace}subspace\left\{  \psi|S\psi
=\psi\right\} \nonumber\\
S\left(  \psi_{1}+i\psi_{2}\right)   &  =\psi_{1}-i\psi_{2},\,\,S=S_{sc}%
S_{in}\nonumber
\end{align}
the position of the dense subspace $H(W)$ inside the total Hilbert space
depends in a subtle way on the interaction through $S_{sc}.$ The domain of
$\Delta^{\frac{1}{2}}$ is now encoded more concretely in terms of the complex
dense space $H(W)$ whose real and imaginary part are vectors in a closed real
subspace $H_{R}(W).$ These real closed subspaces encode the full spatial
aspect of wedge localization. With the help of the graph of the Tomita
involution $S$\ one may even introduce a topology in terms of which the dense
subspace becomes a Hilbert space in its own right\footnote{$H(W)$ with the
$S$-graph norm may be called the thermal Hilbert space, because it offers a
natural description of the (Hawking-Unruh) thermal aspects of the vacuum upon
its restriction to the wedge algebra.}, but all these \textit{spatial}
concepts are still quite remote from the task of characterizing a wedge
\textit{algebra} uniquely in terms of the scattering matrix. The reason is the
following. The algebra-state vector relation $A\longleftrightarrow A\Omega$ is
not universal but changes with the algebra (even within the family of wedges).
In particular the spatial modular theory without additional informations is
too weak to uniquely determine an operator algebra. However the use of
scattering theory is sufficient for obtaining a pure algebraic derivation of
the Bisognano-Wichmann geometric properties for the modular objects without
reference to pointlike field coordinatizations \cite{M}.

Connes has given a criterion \cite{Connes} which allows to obtain from the
spatial modular theory an algebra with the same modular objects. This is
achieved by controlling certain facial properties of subcones of a natural
cone $\mathcal{P}(\mathcal{A}(W))$ associated with $H_{R}(W).$ But one
presently lacks a physical foundation and justification for such a procedure.
Fortunately for our interest in uniqueness, these difficulties can be avoided
if one assumes an additional physically motivated working hypothesis. This is
the crossing property of particle matrix elements (formfactors) of wedge
localized operators. For this we need to remind the reader of a bit of
scattering theory adapted to the algebraic framework.

\section{Uniqueness from KMS-thermality and crossing}

It is well-known \cite{BBS} that any vector $\psi$ which is in the domain of
the positive ``analytically continued'' standard L-boost (\ref{K})
\ $\Delta_{W}^{\frac{1}{2}}$ (which is defined by the use of the functional
calculus of spectral theory on the selfadjoint boost generator) has a unique
relation to an (generally unbounded) operator $F_{\psi,\mathcal{A}(W)}$
affiliated with $\mathcal{A}(W)$ with
\[
F_{\psi,\mathcal{A}(W)}\Omega=\Psi,\,\,F_{\psi,\mathcal{A}(W)}^{\ast}%
\Omega=S_{W}\Psi
\]
But this famous statevector-operator relation depends crucially on the
standard pair ($\mathcal{A}(W),\Omega$). If the same scattering data would
allow for another wedge algebra $\mathcal{B}(W)\neq\mathcal{A}(W),$ the vector
$\Psi\in H(W)$ is associated with another operator $\Psi=F_{\psi
,\mathcal{B}(W)}\Omega$ where $H(W)$ contains all those in- or out- n-particle
vectors which are in the domain of $\Delta_{W}^{\frac{1}{2}}$ which form a
dense set. We have to show that this cannot occur.

Let us assume that we are dealing with a state-vector of the special form
$\Psi=A\Omega,\,A\in\mathcal{A}(W).$ With respect to the $\mathcal{B}(W)$
algebra there exists a unique affiliated densely defined closed operator $F$
with \cite{BBS}
\begin{align}
&  A\Omega=F\Omega\\
&  F\eta\mathcal{B}(W)\nonumber
\end{align}
where in the last line we used the standard notation $\eta$ for a possibly
unbounded \textit{closed operator affiliated with }$\mathcal{B}(W).$ This
forces in particular the inner products with the n-particle out state
vectors\footnote{Here and in the following we omit the smearing with
one-particle wave functions which is necessary to convert improper
(plane-wave) vectors into proper ones.} to be the same
\begin{equation}
^{out}\left\langle p_{n}...p_{1}\left|  A\right|  \Omega\right\rangle
=\,^{out}\left\langle p_{n}...p_{1}\left|  F\right|  \Omega\right\rangle
\label{vac}%
\end{equation}
The physical interpretation would consist in stating that the vectors which
$A$ and $F$ generate from the vacuum do not only possess the same particle
component (as it would be necessary for obtaining the same asymptotic states
with identical normalizations), but their full vacuum polarization clouds (if
the operators are charged, these clouds consists of particle/antiparticle
pairs.) are identical as well. From this we would like to conclude the
equality of their generalized formfactors which then yields the equality of
their matrix-elements between particle states with multi-particle ket vectors
and hence to the identity $A=F.$ For those operators $A\in\mathcal{A}(W)$
which are localized in a double cone $\mathcal{A}(\mathcal{O})\subset
\mathcal{A}(W)$ the LSZ-formalism and on-shell analytic continuation lead to
the crossing symmetry (see appendix)%

\begin{align}
&  ^{out}\left\langle p_{1},p_{2},...p_{l}\left|  A\right|  q_{1}%
,q_{2}...q_{k}\right\rangle ^{in}=\label{cro}\\
&  \underset{p_{c}\rightarrow-p}{a.c.}^{out}\left\langle p_{1},p_{2}%
,...p_{l-1}\left|  A\right|  q_{1},q_{2}...q_{k},(\bar{p}_{c})_{l}%
\right\rangle ^{in}+c.t=\nonumber\\
:  &  ^{out}\left\langle p_{1},p_{2},...p_{l-1}\left|  A\right|  q_{1}%
,q_{2}...q_{k},-\bar{p}_{l}\right\rangle ^{in}+c.t.\nonumber
\end{align}
The notation is the following. The subscript $c$ indicates that the analytic
continuation from momenta $p$ on the positive mass shell to their opposite
values -$p$ is done via the $2\times(n-1)$ dimensional complex mass
shell\footnote{The process of analytic continuation takes place on the complex
mass shell and links the forward and backward parts \cite{Bros} (see
appendix); unfortunately the same terminology is often used in case of an
analytic connection which passes into the complex off-shell region which can
be established much easier.}. The bar on top of the $p$ is only a reminder
that (as demanded by charge conservation) the crossed momentum -$p$ (the end
point of the on-shell analytic continuation) is that of an antiparticle (with
identical Poincar\'{e} characteristics). In this crossing process there arise
a $\delta$-function contraction terms (indicated by $c.t.)$ resulting from a
contraction of $p_{l}$ with one of the $q$ multiplied with a lower particle
formfactors of $A.$ In other words the crossing property consists in naively
crossing $p$'s from outging bras to incoming kets and simultaeously
analytically continuing from $p$ to -$p$ and conjugating the charge carried by
the crossed particle and results in an identity for the connected part of a
formfactor, which in analogy with symmetry identities is often (erroneously)
called ``crossing symmetry''. The crossing of the S-matrix results formally
for the special case $A=\mathbf{1,}$ but it follows slightly different rules
since the unit operator cannot create vacuum polarization but only relates
incoming state vectors to outgoing with a possible different number of real
particles (on-shell creation/annihilation).

In an appendix the reader will be reminded of the ``derivation'' of crossing
in the LSZ-setting of quantum field theory. The necessary on shell analyticity
properties have been derived only in very special cases \cite{Bros}. In the
present context we will simply assume the crossing property. Its deeper
connection with causality, spectral properties and modular theory are
inexorably linked with the existence problem and will be taken up in a
separate paper.

Starting from the matrix element (\ref{vac}), the successive application of
the crossing property (\ref{cro}) allows to obtain connected matrix elements
of $A$ between arbitrary bra-out and ket-in particle states by starting from
the special vacuum polarization situation caused by a local operator
(associated with a compactly localized spacetime region) applied to the
vacuum. From the uniqueness of connected part of the out-in formfactors and
the knowledge of the S-matix one then derives the uniqueness of the in-in
formfactors. If these formfactors are really the matrix elements of a closed
operator, the latter is fixed uniquely.

For free Hermitean fields one can easily that see that the crossing relation
is a direct consequence of the KMS property for the wedge algebra
$\mathcal{A}(W)$ and its affiliated smeared fields. We use the following
notation
\begin{align*}
&  A(f)\equiv\int A(x)\hat{f}(x)d^{4}x=a^{\ast}(f)+h.c.\\
&  \,a^{\ast}(f)=\int a^{\ast}(p(\theta,p_{\perp}))f(p(\theta,p_{\perp}%
))\frac{d\theta}{2}dp_{\perp}\\
&  A(x)=\frac{1}{\left(  2\pi\right)  ^{\frac{3}{2}}}\int\left(  a^{\ast
}(p(\theta,p_{\perp}))e^{ip(\theta,p_{\perp})x}+h.c.\right)  \frac{d\theta}%
{2}dp_{\perp}%
\end{align*}
The $f$ are the wave functions obtained from the W-localized test functions
$\hat{f},$ $supp\hat{f}\in W$ by restricting their Fourier transforms to the
forward mass-shell
\[
p(\theta,p_{\perp})=(m_{eff}ch\theta,m_{eff}sh\theta,p_{\perp}),\,m_{eff}%
=\sqrt{m^{2}+p_{\perp}}%
\]
In the case of the connected 3-pointfunctions we have
\begin{align*}
&  \left(  \Omega,Aa^{\ast}(f_{2})a^{\ast}(f_{1})\Omega\right)  =\left(
\Omega,AA(\hat{f}_{2})A(\hat{f}_{1})\Omega\right)  ^{c},\,A\in\mathcal{A}%
(W),\,A(\hat{f}_{i})\eta\mathcal{A}(W)\\
&  \overset{KMS}{=}\left(  \Omega,A(\hat{f}_{1})\Delta A(\hat{f}_{2}%
)\Omega\right)  ^{c}=\left(  \Delta^{\frac{1}{2}}JA(\hat{f}_{1})\Omega
,Aa^{\ast}(f_{2})\Omega\right)  ^{c}\\
&  \curvearrowright\int\int\left\langle 0\left|  A\right|  p_{2}%
,p_{1}\right\rangle f_{2}(p(\theta_{2},p_{\perp,2})f_{1}(p(\theta_{1}%
,p_{\perp,1}))d\theta_{2}dp_{\perp,2}d\theta_{1}dp_{\perp,1}=\\
&  =\int\int f_{1}((p(\theta_{1}+i\pi,-p_{\perp,1}))\left\langle p_{1}\left|
A\right|  p_{2}\right\rangle ^{c}f_{2}(p(\theta_{2},p_{\perp,2})d\theta
_{2}dp_{\perp,2}d\theta_{1}dp_{\perp,1}%
\end{align*}
By contour shift, the use of analytic properties of matrix elements of
W-localized operators in free field theories and the denseness of the
W-localized wave function spaces one obtains the desired crossing relation
\[
\left\langle 0\left|  A\right|  p_{2},p_{1}\right\rangle =\left\langle
p(\theta_{1}-i\pi,-p_{\perp,1})\left|  A\right|  p_{2}\right\rangle ^{c}%
\equiv\left\langle -p_{1}\left|  A\right|  p_{2}\right\rangle ^{c}%
\]
where the right hand side with the backward momentum in the bra vector is a
shorthand notation for the analytic continuation in $\theta.$ This rather
trivial special illustration has an immediate generalization to an arbitrary
number of particles with arbitrary spin and internal charges\footnote{In that
case the modular $J$ operator involves a charge conjugation and a -1 twist
factor for halfinteger spin.}. Hence in the interaction-free case, crossing
for formfactors (i.e. bilinear forms of wedge-localized operators, in
particular of local composite fields) between multiparticle states follows
from the thermal KMS property for wedges (the Unruh situation) and modular
one-particle properties.

The breakdown of this argument in the presence of interactions (in which case
crossing has the form (\ref{cro})) is related to the fact that generally the
multi-particle in and out state vectors cannot be created by operators
affiliated to the wedge. There exists however a curious exception for those
interacting situations in which wedge-localized so called ``tempered PFG''
exist. These are operators which are localized in $W$ and whose one-fold
application to the vacuum creates a one-particle state without any
vacuum-polarization admixture \cite{BBS}\footnote{Temperedness restrictions
for nonlocal creation anyonic operators actually appeared first in
\cite{Mund}.} (PFGs for sub-wedge localization regions would immediately lead
back to free fields) and have reasonable domain properties with respect to
translations (the temperateness assumption), . The only known realizations for
such situations with Bosons/Fermions are the d=1+1 factorizing models and it
is believed that the family of these models exhaust the possibilities of d=1+1
tempered PFGs.

General PFGs always exist in any QFT, however the temperateness restriction
allows only elastic interactions in d=1+1\footnote{Tempered PFGs in d=1+3
theories are inconsistent with interactions; in fact there exist arguments
that interactions in that case always imply the presence of inelastic
scattering \cite{Aks}.} \cite{BBS}. In fact the only known models with real
particle number conservation are those where the Fourier transforms of the
PFG's $G_{W}(x)$ \cite{S2} fulfill a Zamolodchikov-Faddeev \cite{Zam} algebra
which expresses conservation of particles and individual momenta and which in
the simplest case of a selfconjugate particle reads%

\begin{align}
G_{W}(x)  &  =\frac{1}{\sqrt{2\pi}}\int(e^{-ipx}Z(\theta)+h.c.)d\theta
,\,\,\,p=m(ch\theta,sh\theta)\label{Z}\\
Z(\theta)Z(\theta^{\prime})  &  =S(\theta-\theta^{\prime})Z(\theta^{\prime
})Z(\theta)\nonumber\\
Z(\theta)Z^{\ast}(\theta^{\prime})  &  =S^{-1}(\theta-\theta^{\prime})Z^{\ast
}(\theta^{\prime})Z(\theta)+\delta(\theta-\theta^{\prime})\nonumber
\end{align}

The unitarity of the structure functions $S(\theta)$ is a consequence of the
$^{\ast}$-algebra property of the $Z^{\prime}s$ whereas the KMS property of
the $\,\,G_{W}$-correlation functions with one additional $A\in\mathcal{A}(W)$
follows from the crossing property of the S-matrix; in fact within the setting
of factorizing S-matrices the crossing property is equivalent to that KMS
property which constitutes the thermal characterizes of wedge localization.
The $Z^{\prime}s$ have a simple representation in a bosonic/fermionic Fock
space\footnote{In the presence of one $Z^{\#}$-generator describing several
particles (bound states obeying ``nuclear democracy'') it is simpler to
characterize $Z^{\#}$ by its action on state vectors in the multiparticle Fock
space rather than by its algebraic commutation structure as in (\ref{Z}).}.
Each operator $A$ affiliated with $\mathcal{A}(W)$ has a formal power series expansion%

\begin{equation}
A=\sum\frac{1}{n!}\int_{C}...\int_{C}a_{n}(\theta_{1},...\theta_{n}%
):Z(\theta_{1})...Z(\theta_{n}): \label{series}%
\end{equation}
where $Z(\theta-i\pi)=Z(\theta)^{\ast},$ and each integration path $C$ extends
over the upper and lower part of the rim of the strip. The strip-analyticity
of the coefficient functions $a_{n}$ expresses the wedge-localization of $A.$
The sharpening to double cone localization by the intersection of wedges leads
to meromorphic functions which obey the kinematical pole condition of Smirnov
\cite{Smir}.

Expansions like (\ref{series}) are nothing more than a generating operator for
the formfactors i.e. bilinear forms which fall short of being genuine
operators with domains and closures. They are analogous to the LSZ expansions
of Heisenberg fields into (asymptotic) free fields. For our above uniqueness
argument this is enough, but for a constructive approach this is insufficient.

Although the coefficient functions $S(\theta)$ of the $Z$-algebra turn out to
be the 2-particle scattering matrix, there is no need to know this for the
calculations: absence of real particle production, wedge-localization and the
related KMS property (i.e. spacetime properties) are enough \cite{Lech}.

The case without the temperateness assumption also starts from formfactors
between the dense set of wedge affiliated n-particle-ket-states and the
bra-vacuum (which according to modular theory must be equal for the two
putative theories).The argument then uses crossing symmetry for the successive
movement of particles from the ket to the bra state (and in this way bypasses
domain issues).

The asymptotic states in the $\Delta^{\frac{1}{2}}$ domain have a tensor
product structure in terms of one-particle states. If their wedge
representatives $F_{n}\Omega$ would inherit this factorization structure in
the form $F_{n}\Omega=F_{n-1}G\Omega=F_{n-1}\Omega\times G\Omega$ with $G$
being a PFG, then the sequential crossing from bras to kets would follow from
the KMS formula of wedge localization. But I have not been able to derive such
a factorization from the known domain properties of wedge affiliated
$F^{\prime}s$ $\ $in the general nontemperate case and I doubt that it holds.
This problem is related to the fact that the use of PFG's for the construction
of wedge algebras is restricted to factorizing models \cite{BBS}, in realistic
interacting theories they do not seem to be useful generators of wedge algebras.

\textit{\ }For a recent discussion of how the wedge algebras $\mathcal{A}(W)$
are related to their holographic projections onto the (upper) horizon
$\mathcal{A}(R_{+})$ we refer to \cite{hol}\cite{F-S1}.

\section{Related problems, outlook}

We have seen that by combining modular theory (which among other things gives
mathematical precision to the statevector-operator relation) with the crossing
property (which permits to elevate relations involving vacuum-polarization
formfactors to formfactors describing real particle creation), one obtains a
uniqueness argument for the inverse problem in QFT namely a physically
admissible S-matrix has, in spite of the myriads of interpolating fields, at
most one system of local algebras i.e. at most one
field-coordinatization-independent algebraic QFT.

Whereas the crossing property for interaction-free-theories and the closely
related d=1+1 dimensional factorizing- models follows from the basic
principles of local quantum physics to wedge localization (in fact it is
equivalent to the thermal KMS property), the problem of crossing in the
general setting remains open. In view of the fact that crossing is the deepest
and still mysterious aspect of scattering theory in local quantum physics, its
present use as a working hypothesis for the study of the inverse scattering
problem is at best preliminary state of affairs.

Implementations of the principles of local quantum physics which try to bypass
ultraviolet aspects of the standard approach by placing on-shell objects as
formfactors\footnote{The S-matrix may be viewed as a formfactor of the
identity operator between multi-particle in and out vectors. The mixed
(out-in) formfactors may be converted into (in,in) formfactors i.e. to
sequilinear forms of (singular) operators with repect to the basis of
in-vectors.} into the center touch upon age old problems of particle physics,
which despite the passage of time have lost nothing of their importance.
Beginning as far back as Heisenberg's S-matrix proposal \cite{Heisenberg},
there was the desire to avoid the short distance problems of pointlike field
theory by advocating a pure S-matrix theory. Different from the dual model and
the subsequent string theory which are also in some sense consequences of this
quest, the original intention of the S-matrix approach was to achieve
ultraviolet finiteness by maintaining the principles of QFT, but finding
concepts which favor the use of on-shell quantities instead of integrating
over off-shell vacuum fluctuation of pointlike objects. For a recent review of
these old attempts and the many auxiliary concepts and working hypothesis as
Mandelstam representations, Regge poles, etc. \cite{Olive}\cite{White}.

The present modular setting subjects this old approach to a critical review.
Although there is agreement with its basic premise that on-shell concepts
should play an important role right from the beginning and that there should
exist a different way to introduce interactions than by coupling free fields,
the new approach would not abandon the causality and localization principles
of QFT as was advocated in the old approach. The message would rather be that
one should \textit{avoid} the use of pointlike fields (and their time-ordered
correlation functions) \textit{in intermediate steps of the calculation}. Once
the net of spacetime-indexed operator algebras has been constructed by other
means, there is no harm to use these singular objects (operator-valued
distributions) as generators of these operator algebras and present the
operator-algebra content in terms of field coordinatizations.

There are some fine points in such a program which need to be taken into
account. A useful representative illustration is provided by the
interaction-free theories which correspond to the zero mass ``continuous
spin'' (spin-tower) representations which appear in Wigner's classification of
irreducible Poincar\'{e}-group representations. It is a characteristic
property of this representation that the compactly modular localized subspaces
are empty and the best localized (smallest localization region) nontrivial
subspaces have a semiinfinite string (instead of a point) as the core of their
localization \cite{BGL}\cite{F-S2}. In this case the associated operator
algebras (which are obtained by applying the Weyl functor to the one-particle
localization-subspaces) admit only string-like generators. One expects that
this net of algebras possesses a net of pointlike localizable subalgebras of
observables which consist of string/anti-string such that by the use of the
Doplicher-Roberts reconstruction the original string-localized net appears as
a ``DR field-algebra'' \cite{Haag}. These investigations suggest that the
positive energy restriction guaranties that one never needs to work with
generators which have worse than spacelike cone localization region (with
semiinfinite strings as their core). In the presence of a mass gap this can be
rigorously shown \cite{Bu-Fre2}. For conformally invariant theories the result
in \cite{Joerss} strongly suggest that pointlike generators are always available.

The bootstrap formfactor approach to factorizing models may serve as an
excellent illustration of this new ultraviolet-finite way of thinking about
QFT (see \cite{BFKZ} for a review within the present setting of scattering
theory). Here one bypasses the technical frontiers between
renormalizable/nonrenormalizable interactions by using a different approach to
Lagrangian quantization or causal perturbation of free fields. For d=1+1
factorizing models this is achieved by starting with an algebraic structure
which avoids pointlike fields in favor of operators fulfilling a Z-F algebra
consistent with wedge-like localization. With other words, the system of wedge
algebra $\mathcal{A}(W)$ is constructed \textit{before} any pointlike field
appears on the scene. The next step, namely to get from noncompact wedges to
compact localization regions, consists in the formation of double cone
intersection algebras; this leads to the so-called kinematical pole equation
which relates the lower with the higher formfactors and defines the formfactor
spaces for double-cone localized objects. Whereas the Z--F algebra generators
are PFGs, the double-cone localization leads to the vacuum polarization clouds
in form of expansions with respect to Z-F operators which resemble the LSZ
expansions of Heisenberg operators in terms of incoming free fields
(\ref{series}). The finite size of the spacetime extension of the double-cone
localized \cite{S2} operators shows up in form of a Payley-Wiener asymptotic
behavior of the meromorphic formfactors.

The use of pointlike fields within formfactors (i.e. avoiding correlations)
causes no problems; the pointlike nature is reflected in a polynomial behavior
in certain reduced formfactors \cite{BFKZ}. The point which needs to be
emphasized here is there is no ultraviolet limitation coming from
power-counting and leading to the standard separation into renormalizable and
nonrenormalizable coupling; every admissable factorizing S-Matrix leads to
power bounded formfactors in terms of a few physical parameters which were
already present in the S-matrix. Presently it appears that any S-matrix with
the crossing property and one-particle poles (obtained by applying the fusion
rules to the given 2-particle S-matrix) leads to the formfactors of an
existing QFT. Any potential further restriction on the pole structure must
come from the modular wedge localization requirement. One needs investigations
which go beyond \cite{Lech} in order to clarify this point.

It is comforting to observe that the limitation of Lagrangian fields (which
must have operator dimension near the canonical free field dimension in order
to maintain renormalizability) disappears in this bootstrap-formfactor
setting; in fact any factorizing S-matrix always leads to a renormalizable
theory in the sense of polynomial bounded high energy behavior and a finite
number of physical parameters). So the standard
renormalizable/nonrenormalizable separation according to short distance
behavior becomes void; short distance properties (of what? there is no
preferred Lagrangian field coordinate!) are simply not part of the modular
program and ``nonrenormalizable'' in the Lagrangian sense presumably
corresponds to the triviality of algebras localized in double cone
intersections in the present setting. The remaining question about existence
is whether the double cone intersection algebras are nontrivial in the sense
of formfactors and whether these formfactors are really coming from operators
(or whether the correlation functions exist and have the right properties).

The crucial question is whether a construction based on formfactors and
modular concepts is also conceivable with realistic S-matrices which describe
on-shell particle creation approach. The present uniqueness argument of the
inverse problem suggests that this should indeed be expected. The fact that
the crossing property is an equation for the connected part of the formfactors
and not for the matrix elements themselves complicates its encoding into an
operator approach. There are arguments in favor of existence of an auxiliary
operator formalism for connected formfactors of very special objects. These
attempts at a general on-shell construction require new concepts and will be
presented in a separate paper \cite{constr}. As the standard perturbation
theory, the formfactor approach should admit iterative solution with the tree
approximation as an input. Since such an approach does not require to deal
with correlations between several pointlike fields (i.e. with singular short
distance fluctuations), short distance problems and the related ultraviolet
divergencies do not enter. Naturally one expects that the S-matrix and
formfactors of renormalizable theories in the standard sense to be also
solutions in the new setting.

In particular one expects a clarification of the ghost issue. This is because
the necessity to introduce cohomological BRST ghosts is inexorably related
with controlling the short distance fluctuations in intermediated steps of the
Feynman approach. But since they leave no traces in the physical operator
algebras after the cohomological descend to the physical space (similar to
catalyzers in chemistry \cite{D-S}), there is no place for ghosts in an
approach which bypasses the short distance aspects.

In this way one may hope for a return of Heisenberg's credo that quantum
physics should admit a formulation solely in terms of observables. In fact
local quantum physics provides an excellent illustration of the power of this
idea. The locality structure is such a strong restriction that the structure
of the (neutral) observable subalgebras allows to reconstruct the statistics
(spacelike commutation structure) and the internal symmetry of the
charge-carrying field algebra from its observable ``shadow'' \cite{Haag}. This
is a perfect local analog of Marc Kac's aphorism with which we started this
paper. The uniqueness and constructive existence of the inverse scattering
problem extends this analogy to the global (asymptotic) domain. In quantum
mechanics this is only possible under severe restrictions on the interaction potential.

New ideas for which one has only very special illustrations are sometimes
easier communicated by pointing to the underlying philosophy. It has been
known that without the presence of interaction terms, i.e. for free Wigner
particles with arbitrary spin/helicity, there does indeed exist a modular
approach leading directly to the net of algebras. In this case the S-matrix
input trivializes to the Wigner one-particle representation space and the
wedge localization amounts to a construction of geometrically defined real
subspaces and their intersections \cite{S1}\cite{BGL}. The application of the
Weyl- (or CAR-) functor to these subspaces yields a system of extended
generators of the net of von Neumann algebras.

The standard method which uses pointlike fields is different, although it also
starts from the same Wigner one-particle representations. Pointlike covariant
fields result from intertwiners between the unique (m,s) Wigner
representations and the multitude of so-called covariant representations
\cite{Wein} which form a infinite denumerable set. In this way one obtains
infinitely many covariant fields which live in the same multiparticle tensor
Fock space and which define different field coordinatizations of the same
algebraic net, i.e. the uniqueness of the Wigner (m,s) representation theory
gets lost in the construction of the fields and is recovered by passing to the
net of algebras. Actually this system of fields from different choices of
intertwiners exhausts only the linear part of the system of all possible
pointlike field coordinatizations. The full system is identical to the Wick
polynomials formed from the linear system. The analog system in the case with
interactions is the Borchers class of the theory\footnote{One obtains a very
interesting generalization of this situation for so-called generalized free
fields \cite{D-R}. An analysis of that situation in the present setting of
modular subspaces and Weyl functors could lead to simplifications and new
insights.}. The interacting case also starts with Wigner particle data, this
time one needs in addition an invarant quantitative characterization of their
interaction which is the S-matrix. The latter characterizes the position of
the dense subspace of wedge-localized state vectors within the Fock space of
incoming particles. The remaining problem of how to go from here to the net of
subalgebras is the difficult step which needs new conceptual and mathematical constructs.

I believe that it is fruitful to view the general modular approach as an
extension of Wigner's program. Wigner was the first who succeeded to describe
particles in a completely intrinsic manner without using ``quantization'' of
classical fields. Therefore the program advocated in this paper should be
viewed as a generalization of Wigner's approach to full QFT in the presence of interactions.

\subsection{Appendix: Crossing symmetry with LSZ reduction}

Crossing has been first observed in Feynman perturbation before a formal
derivation was given in the setting of LSZ scattering theory. Its formal
aspects are easily obtained from the LSZ asymptotic convergence
\begin{align}
&  lim_{t\rightarrow\mp\infty}A^{\#}(f_{t})\Phi=A^{\#}(f)_{in,out}%
\Phi,\,\,A^{\#}=A\,\,or\,\,A^{\ast}\\
&  A(f_{t})=\int f_{t}(x)U(x)AU^{\ast}(x)d^{4}x,\,\,A\in\mathcal{A}%
(\mathcal{O})\nonumber\\
f_{t}(x)  &  =\frac{1}{\left(  2\pi\right)  ^{2}}\int e^{i\left(  p_{0}%
-\omega(p)\right)  t-ipx}f(\vec{p})d^{4}x,\,\,\omega(p)=\sqrt{\vec{p}%
^{2}+m^{2}}\nonumber
\end{align}
which can be derived on a dense set of states. This is known to lead to the
well-known reduction formulas \cite{Hepp}\cite{Araki} (leaving out the
smearing with one-particle wavefunctions) which in terms of connected matrix
elements read
\begin{align}
&  ^{out}\left\langle q_{1},q_{2},...q_{m}\left|  F\right|  p_{1}%
,p_{2}...p_{n}\right\rangle ^{in}|_{conn}=\\
&  -i\int\,^{out}\left\langle q_{2},...q_{m}\left|  K_{y}TFA^{\ast}(y)\right|
p_{1},p_{2}...p_{n}\right\rangle ^{in}d^{4}ye^{-iq_{1}y}=\nonumber\\
&  -i\int\,^{out}\left\langle q_{1},q_{2},...q_{m}\left|  K_{y}TFA(y)\right|
p_{2}...p_{n}\right\rangle ^{in}d^{4}ye^{ip_{1}y}\,=\nonumber
\end{align}
Here the time-ordering $T$ between the original operator $F\in\mathcal{A}%
(\mathcal{O})$ and the interpolating Heisenberg field $A(x)$ or $A^{\ast}(x)$
appears in the reduction of a particle from the bra- or ket state. For the
definition of the time ordering of a fixed finitely localized operator $F$ and
a field with variable localization $y$ we may use $TFA(y)=\theta
(-y)FA(y)+\theta(y)A(y)F,$ however as we place the momenta on-shell, the
definition of time ordering for $y$ near $locF$ is irrelevant\footnote{These
on-shell reduction formulas remain valid if one used as interpolating
pointlike fields $A(x)$ the translates of bounded localized operators
\cite{Araki}.}. Each such reduction is accompanied by another disconnected
contribution in which the creation operator of an outgoing particle say
$a_{out}^{\ast}(q_{1})$ changes to an incoming $a_{in}(q_{1})$ acting on the
incoming configuration (and the opposite situation i.e. $a_{in}^{\ast}%
(p_{1})\rightarrow a_{out}(p_{1})$). These terms (which contain formfactors
with one particle less in the bra- and ket- vektors) have been omitted since
they do not contribute to generic nonoverlapping momentum contributions and to
the analytic continuations. Under the assumption that there is an analytic
path from $p\rightarrow-p\,$\ (or $\theta\rightarrow\theta-i\pi$ in the wedge
adapted rapidity parametrization) the comparison between the two expressions
gives the desired crossing symmetry: a particle of momentum p in the ket state
within the connected part of a formfactor is indistinguishable from a bra
antiparticle at momentum -p (here denoted as -\={p}).

In order to obtain that required analytic path on the complex mass-shell of
the $2\rightarrow2$ scattering amplitude it is convenient to pass from time
ordering to retardation
\begin{equation}
TFA(y)=RFA(y)+\left\{  F,A(y)\right\}
\end{equation}
The unordered (anticommutator) term does not have the pole structure on which
the Klein-Gordon operator $K_{y}$ can have a nontrivial on-shell action and
therefore drops out. The application of the JLD spectral representation puts
the p-dependence into the denominator of the integrand of an integral
representation where the construction of the analytic path proceeds in a
completely analog fashion to the derivation of crossing for the S-matrix
\cite{BLOT}\cite{Araki}. Whereas it is fairly easy to find an off-shell
analytic path, the construction of an on-shell path which remains in the
complex mass shell is a significantly more difficult matter \cite{Bros}.

The simplifications of the LSZ formalism resulting from factorizability of
models can be found in an appendix of \cite{BFKZ}.

\end{document}